\documentclass[10pt, conference, compsocconf]{IEEEtran}
\ifCLASSINFOpdf
\else
\fi
\IEEEoverridecommandlockouts
\usepackage{cite}
\usepackage{amsmath,amssymb,amsfonts}
\usepackage{algorithmic}
\usepackage{graphicx}
\usepackage{textcomp}
\def\BibTeX{{\rm B\kern-.05em{\sc i\kern-.025em b}\kern-.08em
    T\kern-.1667em\lower.7ex\hbox{E}\kern-.125emX}}

\begin{document}
%
\title{Data-Mining Research in Education\\
{\footnotesize \textsuperscript{}}
}


\author{\IEEEauthorblockN{Jiechao Cheng\textsuperscript}
\IEEEauthorblockA{International School of Software\\
Wuhan University\\
Wuhan, China\\
$jetrobert19@gmail.com$}
}


%


\maketitle

\begin{abstract}
As an interdisciplinary discipline, data mining (DM) is popular in education area especially when examining students' learning performances. It focuses on analyzing educational related data to develop models for improving learners' learning experiences and enhancing institutional effectiveness. Therefore, DM does help education institutions provide high-quality education for its learners. Applying data mining in education also known as educational data mining (EDM), which enables to better understand how students learn and identify how improve educational outcomes. Present paper is designed to justify the capabilities of data mining approaches in the field of education. The latest trends on EDM research are introduced in this review. Several specific algorithms, methods, applications and gaps in the current literature and future insights are discussed here.
\end{abstract}


%
\IEEEpeerreviewmaketitle

\section{INTRODUNCTION}
One of the biggest challenges that educational institutions facing today is the exponential growth of educational data and how to apply this data to improve the quality of managerial decisions [1]. Education Institutions would like to know, for instance, which students will enroll in particular course programs, and which students will need assistance for graduation. Through the analysis and presentation of data they collected, or data mining, the challenges of these student or learners are able be effectively addressed. Data mining enables organizations to uncover and understand hidden patterns in vast databases by using their current reporting capabilities. And these patterns are then built into data mining models and applied to predict individual behavior and performance with high accuracy. In this way, resources and staff can be allocated by institutions more effectively. Data mining may also, for example, efficiently allocate resources with an accurate estimate of how many students will take action before he or she drops out.

Educational data mining (EDM) is an emerging discipline including but not limited to information retrieval, recommender systems, visual data analytics, social network analysis (SNA), cognitive psychology, psychometrics, and so on. Its methods is often different from those methods from the broader data mining literature. What's more, EDM draws from several reference disciplines including data mining, learning theory, data visualization, machine learning and psychometrics [2]. And this emerging field of EDM examines the unique ways of applying data mining techniques to solve educationally related issues. 

This paper is to synthesize and share various examples by using data mining in education, to support reflection on teaching and learning. The background of EDM is described, then various algorithms that frequently used are briefly presented. Some specific EDM methods found are described. Subsequently, several examples of applications demonstrate how data mining are used to save resources and help teachers and learners. Finally, we conclude the paper.

\section{BACKGROUND OF DATA MINING}
Data mining is an interdisciplinary subfield of computer science [3-5]. Data mining is the analysis step of the "knowledge discovery in databases" process, or KDD [6]. Data mining techniques have their roots in machine learning, artificial intelligence, computer science, and statistics etc. [7]. And data mining is an exploratory process, but it can be used for confirmatory investigations [8]. It is different from other searching and analysis techniques because data mining is highly exploratory, where other analyses are typically problem-driven and confirmatory. Through the combination of an explicit knowledge base, sophisticated analytical skills, and domain knowledge, hidden trends and patterns are able to be uncovered. These trends and patterns form the predictive models that enable to assist organizations with uncovering useful information then guide decision-making [9].

The Cross Industry Standard Process for Data Mining (CRISP-DM) is a cycle process for development and analysis of data mining models [10]. As the demand for data mining increases and more algorithms are created, CRISP-DM ensures practices that everyone can follow, and it gives specific tips and techniques on how to understand business data by deploying a data-mining model. CRISP-DM has six phases including business understanding, data understanding, data preparation, modeling, evaluation, and deployment [10].

\section{BACKGROUND OF EDUCATIONAL DATA MINING}
Educational data mining as a field for solving educationally-related problems, at a high level, it seeks solutions to improve methods for exploring the data, which usually has meaningful hierarchy at multiple levels, in order to discover new insights of how people learn in the context of these settings [11]. For instance, a student's college transcript may contain a temporally ordered list of courses taken by him or her, the grade that the student earned in each course, and information about when the student selected or changed his or her academic major or minor. We might also understand how different individuals engage with or potentially 'game' the EDM system. Taken together, these learning analytics provide much useful information for the design of learning environments.

EDM applies lots of techniques such as Decision Trees, Neural Networks, Naïve Bayes, K-Nearest neighbor, etc. into its examples. Qualitative techniques such as interviews and document analysis are frequently used to support case studies in EDM. EDM impact students directly within areas of course content, course selections, recommender systems and admissions. In addition, applications of specific data mining methods like web mining, classification and multivariate statistics are key techniques used in educationally related data [12]. These approaches can be applied into modeling students’ individual difference and respond to those differences by providing a way, which help improve students’ performance [13].

\section{ALGORITHMS OF DATA MINING}
Data mining relies on disciplines like classification, categorization, estimation, and visualization. Classification assists with identifying associations and clusters, and separates subjects under study. E.g., education institutions can use classification comprehensively to analyze students' characteristics. Categorization applies rule induction algorithms to handle categorical outcomes. Estimation includes predictive functions or likelihood deals with continuous outcome variables. Estimation and classification use unsupervised or supervised modeling techniques. Visualization uses interactive graphs to demonstrate mathematically induced data and scores, and is much more sophisticated than traditional bar charts or pie charts. 

An algorithm is a specific, mathematically driven data mining function, such as a neural network, classification and 
Data mining relies on disciplines like classification, categorization, estimation, and visualization. Classification assists with identifying associations and clusters, and separates subjects under study. E.g., education institutions can use classification comprehensively to analyze students' characteristics. Categorization applies rule induction algorithms to handle categorical outcomes. Estimation includes predictive functions or likelihood deals with continuous outcome variables. Estimation and classification use unsupervised or supervised modeling techniques. Visualization uses interactive graphs to demonstrate mathematically induced data and scores, and is much more sophisticated than traditional bar charts or pie charts. 

An algorithm is a specific, mathematically driven data mining function, such as a neural network, classification and regression tree (C\& RT), or K-means. Data mining techniques including algorithms such as clustering, classification, regression, neural networks, association rules, decision trees, some of them have been applied successfully in the educational area [14]. E.g., methods for hierarchical data mining and longitudinal data modeling have been applied into EDM.

\subsection{Clustering}
The goal of clustering is to find data points that naturally group together, splitting the full data set into clusters sets. By using clustering techniques we are able to further identify dense and sparse regions in object space, and discover overall distribution pattern and correlations among data attributes. Here are some researches about clustering techniques [15-17, 18, 19].

\subsection{Classification}
Classification is used to predict values for some variables. This algorithm frequently employs the decision tree or neural network-based classification algorithms. In classification test, data are used to estimate the accuracy of the classification rules. If the accuracy is acceptable then rules are able to be applied into new data. Some popular classification methods include decision trees, logistic regression (for binary predictions) and support vector machines.

\subsection{Association rule}
Association rules are used to find relations between different items [15]. Back to 1995, the analysis method of association rule was frequently utilized in most studies on educational data mining because of its less extensive expertise while comparing with other methods [20, 21]. However, after the year of 2005, as researchers frequently adopting clustering and classification methods, the trend changed.

\subsection{Regression}
Regression analysis can be applied to model the relationship between independent variables and dependent variables. Independent variables are attributes we already known and response variables are able to predict what we want. But a number of real-world problems are not simply prediction. Therefore, more complex techniques (e.g., decision trees, or neural nets) may be necessary for future prediction. Some popular regression methods within educational data mining include linear regression, neural networks, and support vector machine regression.

\subsection{Neural Networks}
Neural networks have the remarkable ability to obtain meaning from complicated or imprecise data and can be used to extract patterns and detect trends that are too complex to be noticed by humans or computer techniques, they are good at identifying patterns or trends for future forecasting needs. 

\subsection{Decision Trees}
Decision tree is tree-shaped structures, which represents decisions sets. Specific decision tree methods include classification and regression trees and Chi Square Automatic Interaction Detection.

\subsection{Nearest Neighbor Method}
Nearest neighbor method, also called the k-nearest neighbor technique sometimes, classifies each record in the dataset based on a combination of different classes of the k record(s), which is similar to that in a historical dataset (where k >=1).

To choose the appropriate algorithms, researchers need design the data and align it with the desired output. As for small-scale, they can opt for clustering approach since it does not require necessary splitting data in classification.

\section{METHODS OF EDUCATIONAL DATA MINING}
EDM has a large number of powerful methods [2, 22], some of which are widely acknowledged to be universal data mining types (E.g., clustering, prediction, outlier detecting, relationship mining, etc.). However, Discovery with Models and Distillation of Data for Human Judgment are considered more prominent methods recently in educational data mining [14].

\subsection{Clustering}
Clustering approaches had been applied to obtain clear distinction between the clusters. Once a set of clusters has been determined, new instances can be classified by determining the closest cluster. Clustering is able to be applied into grouping similar course materials or grouping students based on their learning behavior patterns [24]. There was an e-learning report was design to forecast the students’ behavior pattern within the data mining approaches [25].

\subsubsection{Prediction}
Prediction can deduce single aspect of the data from combinations of data in other aspects. Classification, regression and density estimation are main types of prediction methods. Prediction has already applied into predicting students' performance [26].

\subsubsection{Relationship Mining}
Relationship mining can identify relationships among variables and encode them in rules for later application. Broadly, there are four different types of relationship mining: association rule mining, sequential pattern mining, correlation mining, and causal data mining, some of them has already utilized into identifying the relationship in patterns of students' behavior, difficulties or mistakes that learners usually encountered with [27].

\subsubsection{Process Mining}
Process mining has been used to extract data from event logs in an information system to form a clear presentation in the overall activities. There are three different subfields of process mining: model discovery, model extension and conformance checking. Process mining is reported able to reflect students' behavior in the sequence of course, grade etc. [28].

\subsection{Text Mining}
Text mining can be used for deriving information with high accuracy from text data and resources. The main contents of the text mining contain text categorization, text clustering, concept/entity extraction sentiment analysis and document summarization etc. Text mining is applied to analyze contents from forums, chats, Webpage or text resources etc. [29].

\subsection{Social Network Analysis}
Social Network Analysis (SNA) is used to measure the relationships among different entities of information, and it is able to analyze the relationships in various tasks [30].

\subsection{Discovery with Models}
In discovery with models, a model is validated via prediction, clustering, or manual knowledge engineering. Key applications of this method include discovering relationships among student behaviors, characteristics and contextual variables [14, 31]. 

\subsection{Distillation of Data for Human Judgment}
Distilled data enables humans to identify well-known patterns for identification, it is also applicable for classifying data features. The goal of this method is to summarize and present the information in a useful, interactive and visually appealing way for understanding the large amounts of education data and supporting decision making [32]. Data is distilled for human judgment in educational data mining for two main purposes: identification and classification.

\section{APPLICATIONS IN EDUCATION}
There are three applications of educational data mining with having received particular attention are discussed here. 

\subsection{Predicting Student Performance}
Lin applied classification and regression trees to predict what types of students would drop out from school, and then return to school later on [33]. The models were able to provide short-term accuracy for predicting which types of students would benefit from student retention programs [33]. Chacon and Spicer et al. developed a system based on data mining helps the institution identify and respond to students at-risk [34]. Their work is highly representative of the discipline, because it follows with a strict data mining process, which is quantitative. The research by Yeats, Reddy and Wheeler found that students who attend writing centers tend to do a good job in their classes [35]. Yu and DiGangi, et al. discovered that east coast students in USA tend to keep enrolled longer than their west coast counterparts do [36].

\subsection{Course Management System}
EDM is often used in course management systems, like Moodle, which contains usage data that includes different activities. García, Romero, Ventura, and de Castro developed a simplified data mining toolkit that operates within the course management system and allows students and their learning users to get data mining information for their courses [37]. This research and application contributions will allow non-technical faculty to engage in educational data mining activities. Instead of traditional static course patterns, data mining can be applied to customize learning activities and adapt the pace for learners to complete courses [38, 39]. It will create significant and optimal learning experiences for each student. Also, Blikstein found different types of programming behaviors in an online course [23]. 

\subsection{Planning and Scheduling}
Researches on mobile learning environments recently suggest that data mining can be applied to help provide personalized contents to different mobile users, despite the differences between mobile devices and conventional PCs. EDM applications will allow non-technical users engage in data mining tools and activities making processing more accessible for all EDM users [39]. There are some examples, including statistical and visualization tools, analyzing social networks and related influence on learning outcomes [27].

\section{CHALLENGES}
As the related technology developed, costs and challenges associated with implementing EDM applications, like storing logged data and managing data systems [19]. Moreover, choosing which data to mine and analyze may also be a challenge. In addition, individual privacy is a continued concern for the application of educational data mining tools. With free, accessible tools in the market, students and learners may be at risk providing information to the learning system. Protecting individual privacy should be considered for the long-tern development of EDM. Moreover, it’s unclear what data displays, visualizations, and visual analysis are most informative and support effective decision making for different stakeholders.

\section{CONCLUSION AND FUTURE INSIGHTS}
Data mining is a powerful analytical tool to enhance decision making and analyzing new patterns and relationships for organizations. And EDM contains techniques including data mining, statistics, machine learning. DM need to analyze data coming from teaching and learning, tests learning theories, and policy decision-making etc. There are a number of opportunities exist in EDM, from an analysis at organizational level to the analysis at individual level. What’s more, EDM is widely used and applied by learners, researchers and teachers, even institutions. 

Recently, there are several studies focus on applying EDM into admissions and enrollment, but we don’t know exactly how institutions using data mining to enhance student learning or improving related educational processes. And results from EDM research are typically obtained from the narrow context of specific educational settings. Therefore, the need for studies to examine in the broader context is necessary. For the overall EDM work to be completed, the urgent need of examining how to widespread the adoption of educational data mining is necessary. Furthermore, research indicates the area of educational data mining is concentrated in western cultures and subsequently, other countries like Asians may not be represented in the related researches and studies. Therefore, applications across multiple contexts should be considered in development of future models [39].

\section*{Acknowledgment}
This work was supported by International School of Software at Wuhan University. The authors want to thank teachers and staffs at International School of Software for the provided computational resources and support, as well as the anonymous reviewers for their insightful comments.



%

\end{document}